\documentclass[aps,prl,twocolumn,groupedaddress,showpacs]{revtex4}

\usepackage{graphicx}

\begin{document}

\title{Shear thickening in molecular liquids characterized by inverse melting}

\author{R.~Angelini$^{1}$, G. Salvi$^{2}$, G.~Ruocco$^{1,2}$}
\affiliation{
$^{1}$ Research center SOFT INFM-CNR c/o Universit\`a di Roma "La Sapienza" I-00185, Roma, Italy.\\
$^{2}$ Dipartimento di Fisica, Universit\`a di Roma "La Sapienza" I-00185, Roma, Italy.\\
}
\begin{abstract}

We studied the rheological behavior of a molecular solution
composed of $\alpha$-cyclodextrin, water and 4-methylpyridine, a
liquid known to undergo inverse melting, at different temperatures
and concentrations. The system shows a marked non-Newtonian
behavior, exhibiting the typical signature of shear thickening.
Specifically, a transition is observed from a Newtonian to a shear
thickening regime at a critical shear rate $\dot{\gamma}_c$. The
value of this critical shear rate as a function of T follows an
Arrhenius behavior $\dot{\gamma}_c(T)=$ B exp$(E_a/K_BT)$, with an
activation energy $E_a$ close to the value of the hydrogen bond
energy of the O-H group of the $\alpha$-CD molecules. We argue
that the increase of viscosity vs shear rate (shear thickening
transition) is due to the formation of hydrogen bonded aggregates
induced by the applied shear field. Finally, we speculate on the
possible interplay between the non-Newtonian rheology and the
inverse melting behavior, proposing a single mechanism to be at
the origin of both phenomena.

\end{abstract}


\maketitle

\date{\today}

The study of complex fluids, like colloidal suspensions,
surfactant solutions, polymeric systems, proteins, glasses and
granular materials has gained during the last decades, the
increasing attention of scientists~\cite{larson,barrat}. Due to
the varieties of interactions that give rise to a rich
phenomenology, these systems represent the prototypes to
understand most of the relevant problems in the physics of soft
matter~\cite{hamley}. Besides the wide interest for fundamental
research, the study of complex fluids is of great relevance in for
practical applications due to the possibility of tailoring
different system properties as for example the mechanical or
thermal properties, or the capacity to solubilize and transport
materials, etc. This class of fluids at variance with simple
liquids is characterized, at molecular level, by the presence of
complex micro-structures which give rise to non-Newtonian
behaviors. Consequently the study of the rheological response of
these systems, provides critical information for their
performances in many industrial applications. In fact, due to
several details of the system like inter-particle interactions,
volume fraction, electrostatic forces, steric repulsion, size and
shape of particles, Brownian motion, hydrodynamic forces etc.,
complex fluids may or may not show the appearance of a dynamical
instability. In this context rheological techniques represent a
widespread method to monitor, investigate and/or predict anomalous
behaviors of complex fluids. Common rheological measurements allow
to investigate phenomena like shear thickening, shear thinning and
shear banding. In general for a Newtonian fluid the viscosity is
shear rate independent while for a non-Newtonian fluid it depends
on the applied shear. For the latter it is intuitive to think that
an applied shear could destroy existent fluid structures giving
rise to a decrease of viscosity (shear thinning), however there
are some fluids which behave in a completely different and
counterintuitive way since when subjected to shear they manifest
an increase of viscosity giving rise to the phenomenon of shear
thickening. This behavior has stimulated the interest around this
phenomenology. Many complex fluids, including dense colloidal
suspensions and dilute surfactant aqueous solutions, exhibit a
shear thickening transition. The continuous development of new
theories and the introduction of modern concepts regarding shear
thickening and the physics of complex fluids and soft matter
promote the realization of targeted experiments and the
implementation of experimental techniques aimed to give
complementary results.
Recently, on the basis of the new knowledge
acquired in the field of soft matter and glass transition a class
of models for shear thickening and ``jamming" in colloidal
suspensions has been proposed~\cite{JR492372005} where ``jamming"
induced by shear indicates an arrested phase. In these models
starting from the mode coupling theory (MCT) of the glass
transition, flow curves and correlation functions are calculated
using memory functions which depend on density, shear stress and
shear rate. In this connection in very recent viscosity
studies~\cite{PRL1030860012009,PRL1000183012008,PRE660604012002}
on suspensions of non-Brownian particles subjected to shear there
is evidence of shear thickening and/or permanent ``jamming". In a
further recent spin model for shear thickening reported by
Sellitto and Kurchan~\cite{PRL952360012005} a generic mechanism
for shear thickening analogous to entropy driven phase reentrance
is proposed. In the latter paper the analogy between
entropy-driven transitions in which systems freeze upon heating
and the transitions in which they jam under the action of
stirring, is exploited. Starting from an Hamiltonian modelling a
liquid that undergoes inverse melting, they introduce a force
field representing stirring and obtain a direct derivation of
shear thickening and glass transition found through MCT by Holmes
et al.~\cite{JR492372005} for an inverse freezing system. This
phenomenon happens when a liquid heated at constant pressure
undergoes a reversible liquid-solid transition, thus generating a
solid with entropy higher than its liquid
counterpart~\cite{Sch04aPRL,Cri05aPRL,Sel06aPRB,Fee03aJCP}.

The theoretical possibility to correlate two apparently
independent system properties, as ``shear thickening" and
``inverse melting", is exciting and promising, however a clear
experimental evidences of the coexistence of these two behaviors
in a single system was -up to now- absent. In this paper we
present the first experimental evidence of shear thickening in a
molecular liquid that is known to undergo inverse
melting~\cite{Pla04aJCP,Tom05aJCP,JCP1261245062007,Ang08aPRE,PM8841092008}.

\begin{figure}[t]
\begin{center}
\includegraphics[width=8.5cm,height=8cm]{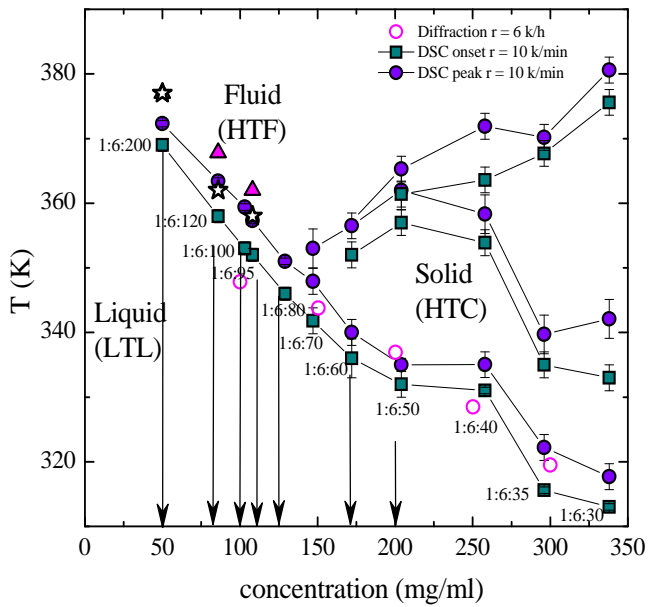}
\caption{Phase diagram of $\alpha$CD-water-4MP solutions from
ref~\cite{Ang08aPRE}. The transition temperatures are plotted as a
function of the concentration of $\alpha$CD in 4MP. The arrows
indicate the concentrations and the high temperature intervals at
which the rheological measurements have been performed.}
\label{fig1}
\end{center}
\end{figure}

We performed rheological measurements as a function of temperature
and concentration on molecular solutions of $\alpha$-cyclodextrin
($\alpha$CD) ($C_{36}H_{60}O_{30}$), water and 4-methyl-piridyne
(4MP) ($C_6H_7N$). The solutions under applied shear exhibit the
phenomenon of shear thickening. In particular we show that the
system goes trough a shear thickening transition at a critical
shear rate $\dot\gamma_c$ which increases with increasing
concentration. It defines two different regimes: at low shear
rates ($\dot\gamma$ $<$ $\dot\gamma_c$ ) the fluid is
Newtonian-like, while at larger shear rates,  $\dot\gamma$  $>$
$\dot\gamma_c$, a transition from the Newtonian behavior to the
shear thickening regime is observed. Viscosity measurements have
been performed at the different concentrations and temperatures
indicated by the arrows in the phase diagram of fig.~\ref{fig1}
with the aim to understand a possible connection between inverse
melting and shear thickening and the nature of the transition
between the two different rheological behavior observed.

The sample was prepared by dispersing $\alpha$CD hydrate in liquid
4MP 98\% purity (both purchased from Aldrich) and deionized water
and stirring for about 4 hours until the suspensions had cleared.
The measurements were performed with a Reostress RS150 rheometer
(Haake) in Couette geometry equipped with two coaxial stainless
steel cylinders. The inner one, the sensor, fixed to a rotor motor
has a diameter of 2 mm (model Z20DIN). The sample was placed in
the 0.85-mm gap between the two cylinders; the temperature of the
outer one was stabilized through a liquid flux cryostat (DC50-K75;
Haake). The rheometer was used in control rate (CR) mode to
measure the shear stress $\sigma$ as a function of shear rate
$\dot{\gamma}$, and hence to determine the shear viscosity
$\eta(\dot{\gamma})$ through the relation
$\sigma(\dot\gamma)=\eta(\dot{\gamma}) \dot{\gamma}$. The
measurements were performed as a function of the shear rate
$\dot{\gamma}$ in the range 0$\div$1400 $s^{-1}$. The sample was
prepared at nine different concentrations of $\alpha$CD, water and
4MP with molar ratios 1:6:x, respectively, with $50 \leq x \leq
200$ as shown in the diagram phase $T-c$ of
fig.~\ref{fig1}~\cite{Ang08aPRE}. The measurement were performed
in the temperature range 259-365 K.

\begin{figure}[t]
\begin{center}
\includegraphics[width=6.5 cm,height=11cm]{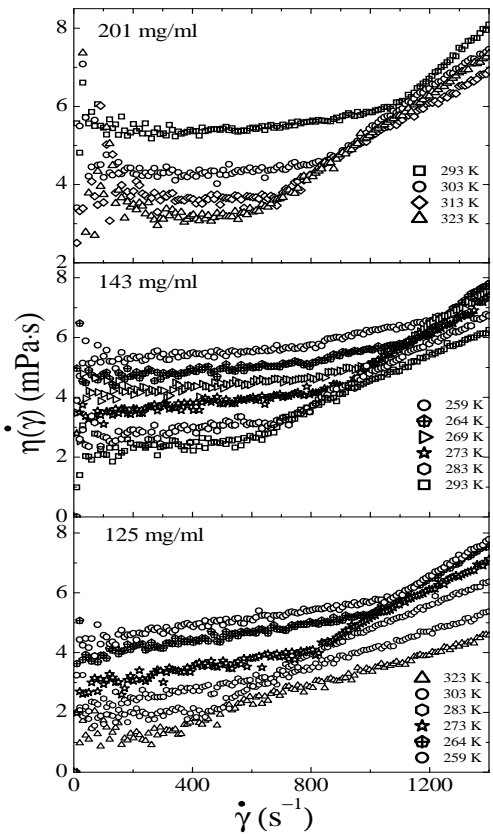}
\caption{Shear viscosity versus shear rate of $\alpha$CD-water-4MP
solutions for three different concentrations of $\alpha$CD in 4MP
(c = 125 mg/ml, c = 143 mg/ml and c = 201 mg/ml) at the indicated
temperatures. } \label{fig2}
\end{center}
\end{figure}

As examples, in fig.~\ref{fig2} the (apparent) shear viscosity
$\eta(\dot\gamma)=\sigma\dot({\gamma})/\dot{\gamma}$ is shown as a
function of the shear rate for few selected temperatures and three
sample concentrations: c = 125 mg/ml, c = 201 mg/ml and c = 143
mg/ml. As exemplified in fig.~\ref{fig2}, the measurements are
characterized by two different regions separated by a critical
shear rate value $\dot{\gamma}_{c}$: at low shear rates, below
$\dot{\gamma}_{c}$, the solutions display Newtonian behavior (the
slight increase is related to spurious inertia effects of the
rheometer as ascertained by performing similar measurements on a
``blank" -water- sample) while at high shear rate, above
$\dot{\gamma}_{c}$, a transition to a clear shear thickening
regime is observed.

\begin{figure}[t]
\begin{center}
\includegraphics[width=7.5cm,height=11cm]{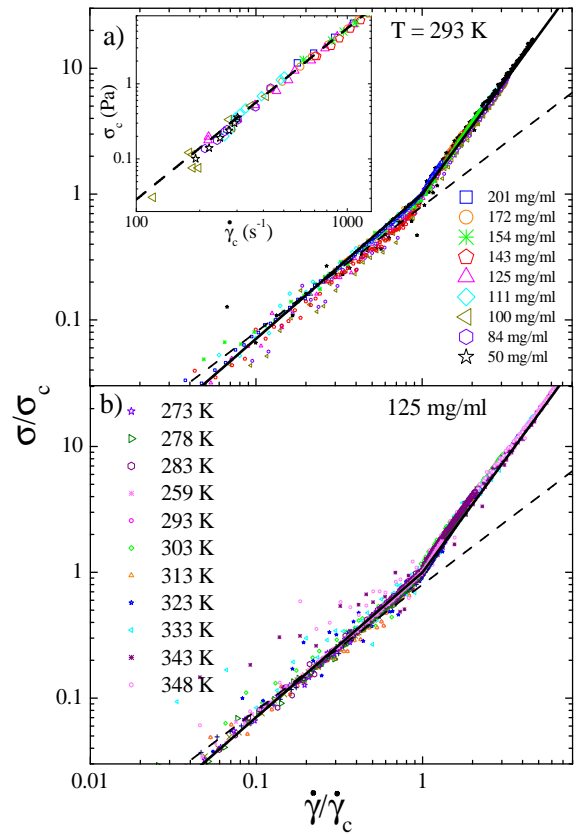}
\caption{ a) Log-log plot of the stress versus shear rate
normalized to their critical values (defined in the text) for
different concentrations at T = 293 K. The power law fits of
experimental data are shown before and after the critical point
(solid line). The Newtonian behaviour (dashed line) is also
plotted for comparison. b) Log-log plot of the stress versus shear
rate normalized to their critical values for different
temperatures at c = 125 mg/ml. In the inset a log-log plot of the
critical stress as a function of critical shear rate is shown at
the indicated temperature for different concentrations of
$\alpha$CD in 4MP together with a power law fit (dashed line).}
\label{fig3}
\end{center}
\end{figure}

From each critical shear rate, $\dot{\gamma}_{c}$, a critical
shear stress value, $\sigma_{c}$, has been obtained as
$\sigma_{c}= \eta(\dot{\gamma}_{c})\dot{\gamma}_{c}$.
Interestingly, as shown in the inset of fig.~\ref{fig3}, a log-log
plot of all the temperatures and concentrations values of
$\sigma_{c}$ vs the corresponding $\dot{\gamma}_{c}$ displays a
single power law behavior $\sigma_{c} \sim
\dot{\gamma}_{c}^{\alpha}$ (the value of the exponent turns out to
be $\alpha=2.28 \pm 0.2$). For each T and c point, the critical
values of viscosity (or the stress) and shear rate were used to
normalize the viscosity curves reported in fig.~\ref{fig2}. The
normalized shear stress, $\sigma/\sigma_{c}$, as a function of
normalized shear rate, $\dot{\gamma}/\dot{\gamma}_c$, are shown in
fig.~\ref{fig3} for selected T and c points. In fig.~\ref{fig3}(a)
the curves at constant T (T = 293 K) at different concentrations
are shown. The data collapse on a unique master curve and, for
$\dot{\gamma}
> \dot{\gamma}_{c}$, they show a common power law behaviour~
\cite{JRheol333291989, JPCM202831032008, PRB731742092006,
PRE740514062006, PRL7820201997} in which the shear stress is
expressed as:
\begin{equation}
{\sigma(\dot{\gamma})\over \sigma_c} = \left ({\dot{\gamma}\over
\dot{\gamma_c}} \right ) ^{n} \label{eq1}
\end{equation}
where $n \sim 1.8$. The fit performed according to the model of
Eq.~\ref{eq1} is plotted (full line) for $\dot{\gamma} >
\dot{\gamma}_{c}$ in fig.~\ref{fig3} and it is compared with the
Newtonian flow corresponding to n=1 and reported as dashed line
for reference. The same behavior of the normalized shear stress vs
normalized shear rate is shown at fix concentration (c = 125
mg/ml) and different temperatures in fig.~\ref{fig3}(b). This
common trend imply that all the measurements scale on a unique
master curve for all the temperatures and concentrations
investigated determining the existence of a single rheological
behavior for the fluid.

\begin{figure}[t]
\begin{center}
\includegraphics[width=8.5cm,height=7.5cm]{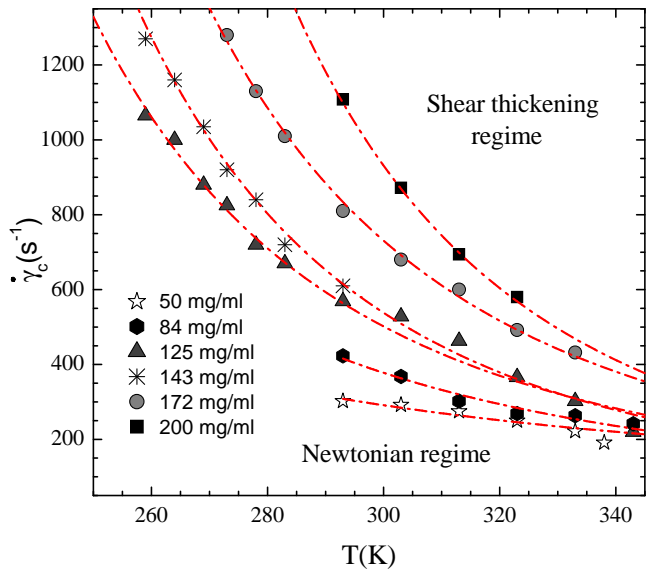}
\caption{Critical shear rate of $\alpha$CD-water-4MP solutions as
a function of temperatures for some of the investigated
concentrations (symbols). The Arrhenius fit of the data (as
discussed in the text) is shown for some of the investigated
concentrations (dash-dot lines). The regions where the solutions
exhibit two different rehological behaviours are displayed.}
\label{fig4}
\end{center}
\end{figure}

\begin{figure}
\begin{center}
\includegraphics[width=8.5cm,height=7.5cm]{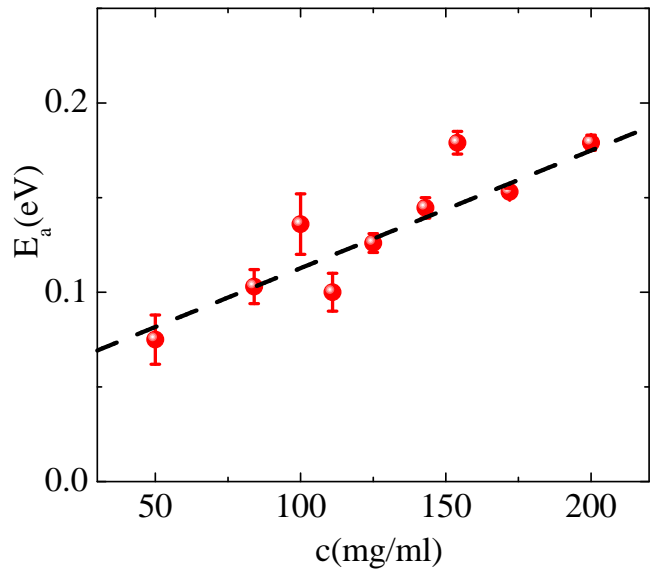}
\caption{Activation energy obtained from the fit reported in
fig.\ref{fig4} as a function of the concentration (symbols). A
linear fit of the data is also shown (dashed line).} \label{fig5}
\end{center}
\end{figure}

In fig.~\ref{fig4} the $\dot{\gamma}_{c}-T$ dynamical phase
diagram is shown, it is evident that the onset of the shear
thickening occurs at larger shear rates as concentration is
increased: the more concentrated the solution, the higher the
shear has to be to observe shear thickening. Moreover the critical
shear rate decreases as the temperature is  increased suggesting
that the shear thickening transition is favored at higher
temperatures. The data follow an Arrhenius behaviour:
\begin{equation}
\dot{\gamma}_c(T)= B e^{\frac{E_a}{K_BT}}\label{eq2}
\end{equation}
where $E_a$ and $K_B$ are respectively the activation energy and
the Boltzmann constant. This behavior seams to resembles the one
found in  solutions of
micelles~\cite{EL416771998,Lang1567551999,EL554322001}, but at
variance with them a decrease of shear rate with temperature is
observed. This apparently counterintuitive phenomenology can be
rationalized by noticing that in an inverse melting system there
is a slowing down of the dynamics at increasing temperature. Let
us consider a characteristic time dictated by the shear flow,
i.~e. $\tau_\gamma= 1/\dot{\gamma}_c$ and the characteristic
relaxation time of the fluid, $\tau_\alpha$, we should expect a
``transition" when these two timescales start to ``interact".
Thus, a decrease of $\dot{\gamma}_c$ implies an increase of
$\tau_\gamma$, consequently of $\tau_\alpha$, and hence a slower
dynamics induced by temperature. The resulting fits with an
Arrhenius law (dash-dot lines in fig.~\ref{fig4}) are in good
agreement with the experimental data. The parameter $E_a$ obtained
from data interpolation is plotted as a function of concentration
in fig.~\ref{fig5}.
It increases linearly with concentration suggesting that a greater
activation energy is required to activate the shear thickening
regime at higher concentrations. In particular we would like to
emphasize that the values found for the activation energy, ranging
between 0.1 and 0.2 eV, are of the same order of magnitude of the
intramolecular hydrogen bonds (HBs) energy, $E_{HB}$, of the
$\alpha$CD molecular O-H group ($E_{HB}$=4.5-4.9 kcal/mol $\equiv$
0.196-0.213 eV)~\cite{RCB5513372006} and of water ($E_{HB} \sim$ 5
kcal/mol) ~\cite{pauling}. This result seems to corroborate the
hypothesis proposed in ref~\cite{PM8841092008} which suggests that
the applied shear rate favours the braking of intramolecular HBs
at a critical value $\dot{\gamma}_c$, which depends either on
concentration and temperature. This in turn could facilitate the
reformation of intermolecular HB with water and $\alpha$CD giving
rise to the formation of more complex aggregations and to an
increase of viscosity. This hypothesis is also reinforced by the
temperature behavior of the critical shear rate of fig.~\ref{fig4}
which decreases at increasing temperature. In fact, the increasing
of T favours the thermal motion of molecules and the breaking of
HBs implying that a smaller $\dot{\gamma}_c$ is required to
activate the shear thickening regime. In addition fig.~\ref{fig4}
suggests that at fixed temperature for higher concentrations of
$\alpha$CD a higher energy is required to break the intramolecular
HBs. Furthermore fig.~\ref{fig5} underline the role played by 4MP
on the shear thickening transition and on the breaking of HBs, it
shows in fact, that the activation energy required for the
transition decreases at decreasing concentration of $\alpha$CD and
departs from the value of the $E_{HB}$ for pure $\alpha$CD
($E_{HB}\sim 0.2 eV$).

These results can also be interpreted considering a time scale
$\tau$ determined by the inverse of the critical shear rate
$\tau_\gamma=1/\dot{\gamma}_c$. When $\dot{\gamma}_c\tau<1$ the
$\alpha$CD molecules subjected to shear have enough time after a
collision to move without remain H-bonded to other molecules. At
increasing shear rate the condition $\dot{\gamma}_c\tau>1$ is
reached and after a collision the $\alpha$CD molecules have no
much time to escape  and they rest bonded to form H-bonded
aggregates which determine an increase of viscosity induced by the
applied shear. This hypothesis of formation of hydrogen bond
clusters had been proposed in absence of shear as a temperature
induced process~\cite{Pla04aJCP}. Here we propose that the same
mechanism is activated with the application of shear and that the
shear rate plays the same role of temperature in favouring the
formation of H-bonded aggregates.

In conclusion in this work we presented rheological measurements
in molecular solutions undergoing inverse melting and we report
the first experimental evidence of shear thickening in such a kind
of systems. The viscosity as a function of the shear rate shows,
at different concentrations and temperatures, two different
behaviors below and above a critical shear rate $\dot{\gamma}_c$:
for $\dot{\gamma}<\dot{\gamma}_c$ the fluid has a Newtonian
behavior while for $\dot{\gamma}>\dot{\gamma}_c$ it enters a shear
thickening regime. The shear thickening transition is observed at
critical shear rates that exhibit an Arrhenius temperature
dependence $\dot{\gamma}_c(T)=$ B exp$(E_a/K_BT)$ where $E_a$ is
the activation energy. $E_a$ has been found very close to the
value of the hydrogen bond energy of the O-H group of the
$\alpha$CD molecules. This finding allows to speculate that the
increase of viscosity and hence the shear thickening behavior
derives from the formation of hydrogen bonded clusters favoured by
the application of shear. This aggregation mechanism activated
here with the application of shear resembles the one suggested to
explain the inverse melting phenomenon activated by
temperature~\cite{Pla04aJCP} . We propose therefore a single
mechanism to be at the origin of both inverse melting induced by
temperature and shear thickening induced by shear field in such a
kind of systems.

\bibliographystyle{apsrev}

\end{document}